\documentclass{ws-ijmpd}
\usepackage[super,compress]{cite}
\usepackage{url}
\usepackage[utf8]{inputenc}
\usepackage[T1]{fontenc}
\usepackage{longtable}
\usepackage{amsmath}
\usepackage{anyfontsize}
\usepackage{amssymb}
\usepackage{amsfonts}
\usepackage{amsbsy}
\usepackage{graphicx}
\usepackage{epsfig}
\usepackage{epstopdf}
\usepackage{ulem}
\usepackage{xcolor}
\usepackage{hyperref}
\definecolor{romared}{RGB}{142,0,28}
\hypersetup{colorlinks=true,
	citecolor=romared,
	linkcolor=romared,
	urlcolor=romared}

\newcommand{\dd}{{\rm d}}

\newcommand{\be}{\begin{equation}}
\newcommand{\ee}{\end{equation}}

\begin{document}

\markboth{Caio F. B. Macedo}
{Scalar modes, spontaneous scalarization and circular null-geodesics of BHs in sGB gravity}

%
\catchline{}{}{}{}{}
%

\title{Scalar modes, spontaneous scalarization and circular null-geodesics of black holes in scalar-Gauss-Bonnet gravity}

\author{Caio F. B. Macedo}

\address{Faculdade de Física, Universidade Federal do Pará, \\
	68721-000, Salinópolis, Pará, Brazil\\
	caiomacedo@ufpa.br}

\maketitle


\begin{abstract}
In general relativity, astrophysical black holes are simple objects, being described by just their mass and spin. 
These simple solutions are not exclusive to general relativity, as they also appear in theories that allow for an extra scalar degree of freedom.
Recently, it was shown that some theories which couple a scalar field with the Gauss-Bonnet invariant can have the same classic general relativity black hole solutions as well as hairy black holes. 
These scalarized solutions can be stable, having an additional ``charge'' term that has an impact on the gravitational-wave emission by black hole binaries. In this paper, we overview black hole solutions in scalar-Gauss-Bonnet gravity, considering self-interacting terms for the scalar field. We present the mode analysis for the mono and dipolar perturbations around the Schwarzschild black hole in scalar-Gauss-Bonnet, showing the transition between stable and unstable solutions. We also present the time-evolution of scalar Gaussian wave packets, analyzing the impact of the scalar-Gauss-Bonnet term in their evolution. We then present some scalarized solutions, showing that nonlinear coupling functions and self-interacting terms can stabilize them. Finally, we compute the light-ring frequency and the Lyapunov exponent, which possibly estimate the black hole quasinormal modes in the eikonal limit.
\end{abstract}


\ccode{PACS numbers:}


\section{Introduction}	

In the past century, black holes (BHs) became a paradigm in physics. More and more evidence amount to the fact that these objects harbor the center of all galaxies, can form binaries with stellar companions, and so on. They became a crucial aspect in describing the structure of the galaxies and the Universe, spanning over a variety of mass-ranges. At the same time, their features pose some of the main reasons why general relativity (GR) should be modified/replaced by an augmented theory of gravity. 
 
One of the main tools to scrutinize higher-curvature regions around compact objects is the detection of gravitational waves (GWs), such as the ones detected by the LIGO-VIRGO collaborations~\cite{Abbott:2016blz,Abbott:2018lct}. By February 2020, we already over 50 possible candidates for gravitational wave sources, most of them from compact binaries. These binaries can be used to test infrared modifications of gravity, especially in the neutron star (NS) binary cases\cite{Sakstein:2017xjx,Ezquiaga:2017ekz,Creminelli:2017sry,Baker:2017hug,Dima:2017pwp,Crisostomi:2017lbg,Langlois:2017dyl}. However, ultraviolet (UV) modifications are more difficult to test. BH-BH binary system seems to be much more common to be detected and beyond GR effects in these are more difficult to probe due, for instance, to numerical challenges, no-hair theorems\cite{Berti:2015itd,Sotiriou:2015lxa,Herdeiro:2015waa}, among others. However, more and more data points towards a majority of BH-BH data, which shows the necessity of exploring theories that present UV modifications of GR.

Among the different candidates that can replace GR, it is crucial to investigate beyond GR theories that have UV modifications and violate the no-hair theorems. An important class that was recently discovered is the one that presents the so-called \textit{BH spontaneous scalarization}\cite{Silva:2017uqg,Doneva:2017bvd,Antoniou:2017acq}. Spontaneous scalarization is a phenomenum by which an initially GR solution spontaneous grows hair through a tachyonic instability. The instability is possible because the additional terms modify the scalar field perturbation inducing an effective negative mass-squared. Spontaneous scalarization was long known to exist for NSs in scalar-tensor theories, where matter terms introduce the effective mass\cite{Damour:1993hw,EspositoFarese:2004cc}. In the BH case, the effective mass is induced by purely spacetime quantities coupled to the scalar field.

The main point for scalarization is to introduce modifications in GR that couple the Gauss-Bonnet term ${\cal G}=R^2-4R_{ab}R^{ab}+R_{abcd}R^{abcd}$ with the scalar field $\varphi$, in a specific manner. Here we explore the action\footnote{Notice that different authors use different quantities for the same theory. We point the reader to Ref.~[\refcite{Macedo:2019sem}] for a discussion on this subject.}
\begin{equation}
S = \frac{1}{2}
\int {\rm d}^4x
\sqrt{-g}
\left[
R - \frac{1}{2}\nabla_a\varphi\nabla^a\varphi - V(\varphi) + f(\varphi){\cal G}
\right]\,,
\label{eq:action}
\end{equation}
where $R$ is the Ricci scalar, $R_{ab}$ the Ricci tensor, and $R_{abcd}$ the Riemann tensor.  $V$ is the self-interacting scalar potential, being
\begin{equation}
V(\varphi)=\frac{1}{2}(\mu^2\varphi^2+\lambda\varphi^4)\,,
\label{eq:potential}
\end{equation}
and $f$ a coupling function, being
\begin{equation}
f(\varphi)=\frac{1}{8}\left(\eta\varphi^2+\frac{1}{2}\zeta\varphi^4\right).
\label{eq:coupling_fun}
\end{equation}
We are using units such that $8\pi G=c=\hbar=1$. The coupling function needs to be at least quadratic in the field to satisfy the condition for the tachyonic instability to occur, 
hence allowing for the scalarization of the Schwarzschild BH~\cite{Silva:2017uqg}.  We shall explore this in more detail in Sec.~\ref{sec:schw}. The scalar field potential is necessary 
complete the UV corrections from the point of view of effective field theory picture. Therefore, we added terms in the potential that still presereve the $\mathbb{Z}_{2}$ symmetry of the 
theory. Note that we only include terms in curvature-scalar coupling that allows for spontaneous scalarization\cite{Macedo:2019sem}.

In this paper, we overview some aspects of sGB gravity that exhibits spontanous scalarization for BHs, by using the theory described above. We complement some of the results presented in Refs.~[\refcite{Silva:2018qhn,Macedo:2019sem}]. The remainder of this paper is organized as follows. In Sec.~\ref{sec:eqs} we present the background field equations, the boundary conditions for BHs and the equations describing radial perturbations. In Sec.~\ref{sec:schw} we extend the stability analysis of the Schwarzschild spacetime by computing the modes and looking into the transition between stable and unstable configurations.  We also look into the interaction of Schwarzschild BHs with scalar Gaussian wave packets. This enable us to understand how scalarized solutions might form in physical scenarios. In Sec.~\ref{sec:scal} we present the scalarized solutions, the light-ring frequency and the respective Lyapunov exponent. Null-geodesics are important to understand the eikonal limit of the quasinormal modes (QNMs) of BHs~\cite{Cardoso:2008bp}. Finally, in Sec.~\ref{sec:final} we present our final remarks.

\section{Equations of motions, boundary conditions and radial perturbations}\label{sec:eqs}

The modified Einstein equations and the scalar field equation can be obtained by extremizing the 
action \eqref{eq:action} with respect to the metric
and the scalar field, respectively. We have
\begin{align}
G_{ab}&=T_{ab}^{\varphi}-\frac{1}{2}{\cal K}_{ab},
\label{eq:einstein_eq}
\\
\Box\varphi&=V_{,\varphi} - f_{,\varphi}{\cal G},
\label{eq:scalar_eq}
\end{align}
where $G_{ab}$ is the Einstein tensor, and
\begin{align}
T_{ab}^\varphi&=\frac{1}{2}\partial_a\varphi\partial_b\varphi-\frac{1}{2}g_{ab}\left[\frac{1}{2}
(\partial_c\varphi)^2+V(\varphi)\right],\\
{\cal K}_{ab}&=2g_{c(a}g_{b)d}\epsilon^{edjg}\nabla_h\left[^*{R^{ch}}_{jg}f'\nabla_e\varphi\right],
\end{align}
with $^*{R^{ab}}_{cd}=\epsilon^{abef}R_{efcd}$.

We focus on static, spherically symmetric BHs. The line element and the scalar field read
\begin{align}
\dd s^2&=-A (r) \dd t^2+B(r)^{-1}\dd r^2+r^2\dd\Omega^2,\label{eq:metric}\\
\varphi&=\varphi_0(r),
\label{eq:scalar_f}
\end{align}
where $\dd\Omega = \dd \theta^2 + \sin^2\theta\, \dd \phi^2$ is the line element on a 2-sphere. The field equations can be obtained by substituting
Eqs.~\eqref{eq:metric} and \eqref{eq:scalar_f} into Eqs.~\eqref{eq:einstein_eq} and~\eqref{eq:scalar_eq}. 
The resulting equations to be solved numerically are
\begin{align}
\label{eq:eintt}(t,t): \qquad &B \left\{ \left(\varphi_{0}''\right)^2 \left[16 (B-1) f_{,\varphi_{0}\varphi_{0}} -r^2\right]+16
(B-1) \varphi_{0}'' f_{,\varphi_{0}} -4\right\}\nonumber\\
&-4 B' \left[2 (1-3 B) \varphi_{0}' f_{,\varphi_{0}}+r\right]-r^2 V(\varphi_{0} )+4=0, \\
\label{eq:einrr}(r,r): \qquad &\frac{1}{4} A \left\{ B \left[4-r^2 \left(\varphi_{0}''\right)^2\right]+r^2 V(\varphi_{0}
)-4\right\}\nonumber\\
&+B A' \left[2 (1-3 B) \varphi_{0}' f_{,\varphi_{0}} + r\right]=0, \\
\label{eq:einqq}(\theta,\theta): \qquad &-r B A'' \left(r-4 B \varphi_{0}' f_{,\varphi_{0}}\right)+4 r B^2
A' \varphi_{0}'' f_{,\varphi_{0}}
-\frac{1}{2} r^2 A \left[B
\varphi_{0}'^2+V(\varphi_{0} )\right] \nonumber\\
& + \frac{1}{2A}\left[r B A'^{2}
\left(r-4 B \varphi_{0}' f_{,\varphi_{0}} \right)\right] -r A B' \\
&+ A' \left\{r B
\left[4 B \varphi_{0}'^2 f_{,\varphi_{0}\varphi_{0}} - 1\right]-\frac{1}{2} r
B' \left(r-12 B \varphi_{0}' f_{,\varphi_{0}} \right)\right\}=0\,,
\end{align}
where a prime indicates derivative with respect to $r$. The equation for the background scalar field is
\begin{align}
&A B \varphi_{0}'' +\frac{1}{2} \varphi_{0}'
\left[BA' +A \left(\frac{4 B}{r}+B'\right)\right]+\frac{4}{r^2}\left[(B-1)
B  A'' f_{,\varphi_{0}}  \right]\nonumber \\
& -\frac{1}{2 r^2 A} \left\{4 A' \left[(B-1)
B A' + A (1-3 B) B'\right] f_{,\varphi_{0}}+r^2 A^2 V_{,\varphi_{0}}\right\}=0.
\label{eq:scalar_back}
\end{align}

The above equations need be supplemented with boundary conditions. Since we are describing BHs in the standard Schwarzschild coordinates, we have
\begin{align}
A(r\approx r_h) &\approx a_1(r-r_h)+{\cal O}[(r-r_h)^2],\label{eq:metric_cond}\\
B(r\approx r_h) &\approx b_1(r-r_h)+{\cal O}[(r-r_h)^2],\\
\varphi_0(r\approx r_h)&\approx \varphi_{0h}+{\cal O}[(r-r_h)],
\end{align}
with $r=r_h$ describing the event horizon position. By plugging these into the differential equations \eqref{eq:eintt}--\eqref{eq:scalar_back}, we obtain the following condition for first derivative of the scalar field at the horizon
\begin{equation}
\left.\frac{\dd \varphi_{0}}{\dd r}\right|_{r=r_{\rm h}}=a^{-1}\left(b+c\sqrt{\Delta}\right),
\label{eq:phi_der}
\end{equation}
where $a,\,b,$ and $c$ are quantities defined in terms of $r_h$, $\varphi_{0h}$, which can be seen explicitly in the Appendix of Ref.~[\refcite{Macedo:2019sem}]. We note that for the solutions to be physical, the discriminant $\Delta$ must be positive, such that the first derivative of the scalar field is real. Its explicit form is given by
\begin{align}
\Delta&=1-6\frac{\varphi_{0h}}{r_h^4}\left(\eta+\zeta\varphi_{0h}^2\right)^2\Bigg\{1-\nonumber\\
&\quad-\frac{1}{2}\varphi_{0h}^2\left(\eta+\zeta\varphi_{0h}^2\right)\left(\mu^2+2\lambda\varphi_{0h}^2\right)\nonumber\\
&\quad-\frac{r_h^2}{6}\varphi_{0h}^2\left(\mu^2+\lambda\varphi_{0h}^2\right)\left[1
+\frac{1}{16r_h^2}\left(\eta\varphi_{0h}+\zeta\varphi_{0h}^3\right)^2\times\right.\nonumber\\
&\quad\left.\left.\times\left(-\frac{24}{r_h^2}+\mu^2\varphi_{0h}^2
+\lambda\varphi_{0h}^4\right)\right]\right\}\,.\label{eq:deltaexpr}
\end{align}
Therefore, $\Delta >0$ is a condition for the BH solution to exist. 

By expanding the field equations for large $r$, requiring asymptotic flatness, we obtain
\begin{align}
A(r\gg r_{\rm h})&\simeq 1-{2M}/{r}\,,
\label{eq:metric_inf}\\
B(r\gg r_{\rm h})&\simeq 1-{2M}/{r}\,,
\\
\varphi_0(r\gg r_{\rm h})&\simeq {Q\,e^{-\mu r}}/{r}\label{eq:scalar_inf}\,,
\end{align}
where $M$ is the ADM mass, $Q$ is an integration constant (which we shall call scalar charge), and we have set the cosmological value of the scalar field to
zero. In general, we start by integrating the field equations from the numerical event horizon for a given $\varphi_{0h}$, increasing $r$, matching with the asymptotic conditions given by Eq.~\eqref{eq:metric_inf}--\eqref{eq:scalar_inf}, extracting the BH mass and the scalar charge $(M,Q)$. The asymptotic behavior defines possible values of the scalar field at the horizon.

The standard Schwarzschild spacetime is a solution of the above equations of motion. It corresponds to the case where the scalar field vanishes identically. However, the dynamical response of the Schwarzschild BHs in sGB theories can be very different from the ones from GR. For instance, new polarizations of modes may appear, such as breathing and longitudinal modes, giving a possible way to distinguish the theories\cite{Will:2014kxa}. Additional modes will be partly explored in Sec.~\ref{sec:schw}.

Radial stability analysis can be performed by considering the following expressions for the scalar field and metric
\begin{align}
\varphi&=\varphi_0(r)+\varepsilon\frac{\varphi_1(t,r)}{r},\label{eq:field_pert}\\
\dd s^2&=[A(r)+\varepsilon F_t(t,r)]\dd t^2+[B(r)^{-1}+\varepsilon F_r(t,r)]\dd r^2+r^2\dd \Omega^2,\label{eq:metric_pert}
\end{align}
where $\varepsilon$ is a small bookkeeping parameter, and $A(r)$ and $B(r)$ and the background metric functions. By plugging these into the field equations \eqref{eq:einstein_eq} and \eqref{eq:scalar_eq} and expanding the differential equations up to first order in $\varepsilon$, taking into account the background field equations, one can show that the system can be reduced to a single second order equation describing $\varphi_1$ (see, e.g., Ref.~[\refcite{Torii:1998gm}] for a detailed description in a similar setup). This equation can be written as
\begin{equation}
h(r)\frac{\partial^2\varphi_1}{\partial t^2}-\frac{\partial^2\varphi_1}{\partial r^2}+k(r)
\frac{\partial\varphi_1}{\partial r}+p(r)\varphi_1=0,
\label{eq:per_scalar}
\end{equation}
where the coefficients
$(h,\,k,\,p)$ depend only on the background quantities and on $r$\cite{Kanti:1995vq,Blazquez-Salcedo:2018jnn,Silva:2018qhn,Macedo:2019sem}. 

For Schwarzschild BHs in sGB, due to the form of the equations, it is sufficient to consider a linear approximation for the scalar field $\varphi$, keeping the metric functions for the background untouched. Through this procedure, we obtain the same equation by a simpler procedure. We shall use this to investigate the dominant (scalar) quasinormal modes as well the linear response to intial data of Gaussian wave packets. Thus, equation describing scalar perturbations of Schwarzschild BH in sGB is given by\cite{Silva:2018qhn}
\begin{equation}
\frac{\partial^2 \varphi_1}{\partial r_{\ast}^2}-\frac{\partial^2\varphi_1}{\partial t^2}
- V_{\rm eff}(r) \,\varphi_1 = 0\,,
\label{eq:eom_perturbation}
\end{equation}
where $r_{\ast}=r+2M \ln(r/2M-1)$ is the tortoise coordinate in the Schwarzschild spacetime, and the effective potential for the perturbations is given by
\begin{equation}
V_{\rm eff} \equiv
\left(1 - \frac{2M}{r} \right)
\left[
\frac{\ell(\ell + 1)}{r^2} + \frac{2M}{r^3}
- \frac{12 M^2\eta}{r^6}+\mu^2
\right]\,.
\label{eq:veff}
\end{equation}
Note that we kept the angular quantum number $\ell$, for generality. We will use equation \eqref{eq:eom_perturbation} to analyze the stability of the Schwarzschild spacetime in sGB.

In what follows, we shall scale our parameters by using the length-scale introduced by $\eta^{1/2}$. We denote the scaled quantities by a hat, e.g.,
\begin{align}
	\hat{M}\equiv M/\eta^{1/2},~	\hat{Q}\equiv Q/\eta^{1/2},~\hat\mu\equiv \mu\eta^{1/2},
\end{align}
and so on. Note that the GR equations for scalar perturbations in Schwarzschild BH should be recovered in the limit $\hat M\gg 1$ (or equivalently $M\gg \eta^{1/2}$).

\section{Stability of Schwarzschild spacetime in scalar-Gauss-Bonnet gravity}\label{sec:schw}

Schwarzschild spacetime is a possible solution of the equations of motion \eqref{eq:eintt}--\eqref{eq:scalar_back} describing spherically symmetric solutions. However, because of the new scalar degree of freedom, the theory now possess new types of modes associated with mono and dipolar emission. Here we analyze the stability of such BHs against spherically symmetric perturbations. Here, for simplicity, we shall focus on the case $\mu=0$, but we expect that the discussion holds qualitatively for massive scalar\footnote{Note, however, that quasibound modes will appear in the massive scalar case.}.

\subsection{Mode analysis}

As noted in Refs.~[\refcite{Silva:2018qhn,Macedo:2019sem,Blazquez-Salcedo:2018jnn}], for the case of Schwarzschild BHs in sGB, there is a critical value for the BH mass $\hat M_c$ bellow which the BH becomes unstable. This unstable mode is purely imaginary, due to the form of the different equations and the boundary conditions. Here we elucidate the transition between stable and unstable configurations, computing for the first time the mono and dipolar scalar QNMs in the region where the Schwarzschild BH are stable.

In order to compute the QNMs, we need to supplement Eq.~\eqref{eq:eom_perturbation} with boundary conditions. QNMs corresponds to purely ingoing waves at the BH event horizon and outgoing waves at infinity. 
By factorizing the time dependent part as $\varphi_1(t,r_\ast)=e^{-i\omega t}\varphi_1(r_\ast)$, the boundary conditions corresponds to
\begin{equation}
\varphi_1(r_\ast)\sim\left\{
\begin{array}{ll}
e^{i \omega r_\ast}, & r_\ast\to\infty ~(r\to\infty)\\
e^{-i \omega r_\ast}, & r_\ast\to-\infty ~(r\to 2 M).
\end{array}\right.\label{eq:bc_pert}
\end{equation}
These boundary conditions generate a boundary value problem whose eigenvalues are the quasinormal mode frequencies of the BH.

There are many methods to find the QNMs of compact objects\cite{Nollert:1999ji,Berti:2007dg,Konoplya:2011qq,Pani:2013pma,Macedo:2016wgh}. Here we explore the direct integration (DI) method~\cite{Chandrasekhar:1975zza} and the continued fraction (CF) one~\cite{Leaver:1985ax,Nollert:1993zz}. The DI essentially consists in integrating the different equation from the horizon and from infinity, matching the two solutions at some intermediate point demanding their Wronskian to vanish. The CF method replaces the differential equation by a recurrence relation, factorizing the divergence of the the spatial equation. In this sense, CF method is more efficient, because we just need to solve a system of linear equations.

As mentioned above, the DI method consist in integrating the differential equation from both the horizon, obtaining $\varphi_{1,-}$ and infinity, obtaining $\varphi_{1,+}$, matching at some intermediate point $r_{\ast m}$. At this point, we require the Wronskian of the two solution to be zero, i.e.,
\be
W(\varphi_{1,+},\varphi_{1,-})=\left[\varphi_{1,+}\varphi_{1,-}'-\varphi_{1,+}'\varphi_{1,-}\right]_{r_{\ast}=r_{\ast m}}=0.
\label{eq:wron}
\ee
Eq.~\eqref{eq:wron} is satisfied if $\omega$ is the QNM frequency of the BH.

For the CF method, we use the following expression for the scalar field perturbation
\be
\varphi_1(r)=\left(\frac{r}{2M}-1\right)^{-2iM\omega}\left(\frac{r}{2M}\right)^{4iM\omega}e^{i\omega r} \sum_{n=0}^{\infty} a_n (1-2M/r)^{n},
\ee
which substituting into the radial part of the scalar field equation 
[once again, considering a time-dependence of the form $\varphi_1(t,r_\ast)=e^{-i\omega t}\varphi_1(r_\ast)$], leads to the six-term recurrence relation
\begin{equation}
\begin{array}{lr}
\alpha_0 a_{1}+\beta_0 a_0=0,&n=0\\
\alpha_1 a_{2}+\beta_1 a_1 +\gamma_1 a_{0}=0,&n=1\\
\alpha_2 a_{3}+\beta_2 a_2 +\gamma_2 a_{1}+\delta_{2}a_{0}=0,&n=2\\
\alpha_3 a_{4}+\beta_3 a_3 +\gamma_3 a_{2}+\delta_{3}a_{1}+\sigma_3 a_{0}=0,&n=3\\
\alpha_n a_{n+1}+\beta_n a_n +\gamma_n a_{n-1}+\delta_{n}a_{n-2}+\sigma_n a_{n-3}+\theta_n a_{n-4}=0,&n>3,
\end{array}
\end{equation}
where
\begin{align}
\alpha_n&=4 M^2 (n+1) (-4 i M \omega +n+1),\\
\beta_n&=3 \eta -4 M^2 \left(\ell^2+\ell+2 n (n+1)+1\right)+128 M^4 \omega ^2+32 i M^3 (2 n \omega +\omega ),\\
\gamma_n&=-4 \left(3 \eta -M^2 (n-4 i M \omega )^2\right),\\
\delta_n&=18 \eta,~~ \sigma_n=-12 \eta, ~~ \theta_n=3 \eta.
\end{align}
The above equation can be reduced to a three-term recurrence relation by multiple Gaussian elimination steps\cite{Pani:2013pma}. This then can be solved by the standard methods to find the QNMs.

The real part of the frequencies dictate the oscillation spectra of the perturbations, while the imaginary tells us about the decaying or growth of the mode. If the imaginary part of the mode is negative, we have that $e^{-i\omega t}$ decays in time, while a positive imaginary part means an exponential growth (unstable mode). By analyzing the boundary conditions \eqref{eq:bc_pert}, we can see that when the mode in unstable, the radial part of the perturbation identically vanishes at the boundaries. Because the differential equation is real, this translates into a boundary value problem for real $\omega^2$. If $\omega^2<0$, we have purely imaginary unstable modes. 

\begin{figure}
	\includegraphics[width=1\linewidth]{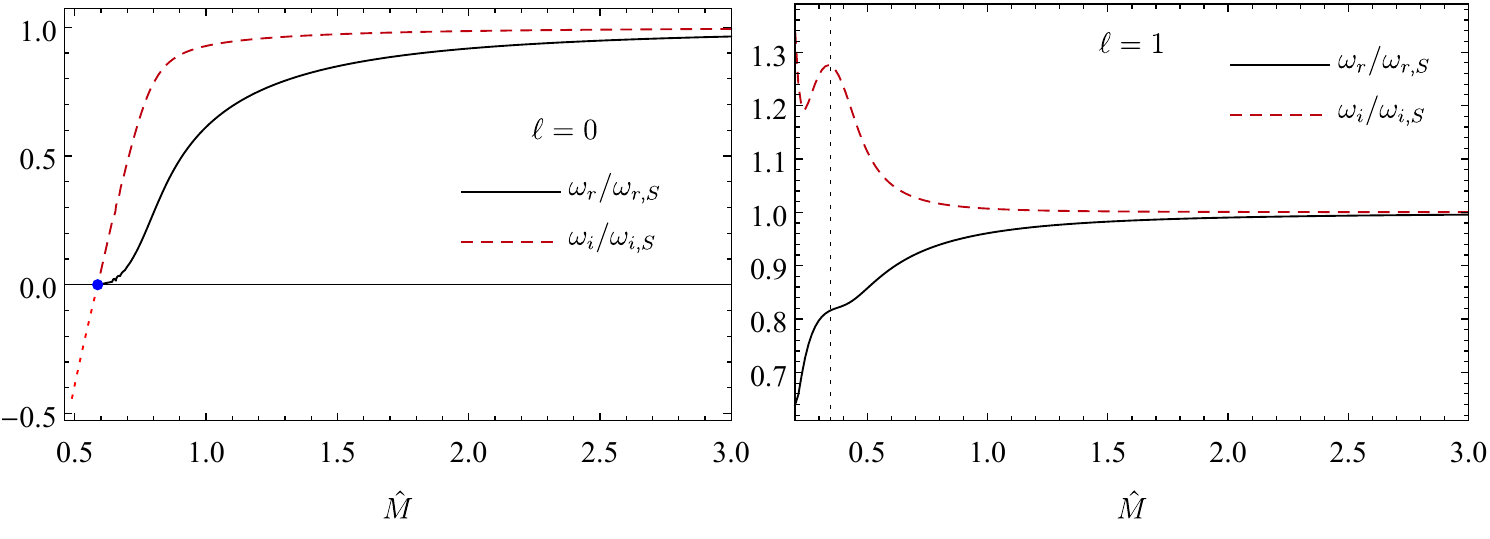}
	\caption{Real and imaginary part of the fundamental mode, normalized by the GR quantities. Note that for $\hat M\gg 1$ (Schwarzschild limit), the curves goes to $1$, as they should. \textit{Left panel}. For the $\ell=0$ case we see that the fundamental mode becomes unstable below a certain critical mass, given by $\hat M_c\approx 0.587$. \textit{Right panel:} The $\ell=1$, although a critical mass for the dipolar mode also exists (represented by a vertical dotted line), the fundamental mode seems to be stable for smaller masses.}
	\label{fig:l_mod}
\end{figure}

The purely imaginary unstable modes in sGB have been investigated at some extension in recent works\cite{Blazquez-Salcedo:2018jnn,Silva:2018qhn,Macedo:2019sem}. Essentially, they exist in a regime of mass $\hat M<\hat M_c\approx 0.587$. Note, however, that for $\hat M>\hat M_c$ stable QNMs exist. The fundamental modes for  $\ell=0$ are shown in the left panel of Fig.~\ref{fig:l_mod}. Note that when $\hat M \to \hat M_c$ from the right, both the imaginary and real part goes to zero, which is expected since the purely imaginary mode arise for  $\hat M<\hat M_c$, and a nonoscilatory scalar field profile exists at $\hat M=\hat M_c$~\cite{Silva:2014fca}. This characteristic tell us that in physical process of BHs near that mass, long-living modes should not be expected because the real part of the frequency goes to zero much faster than the imaginary one. This was also observed in Einstein-dilaton-Gauss-Bonnet gravity~\cite{Benkel:2016kcq}, and we shall investigate this for sGB in the next subsection, where we consider time-evolution of signals. Therefore, any oscillations near the critical mass should be due to the overtones. 

In the right-panel of Fig.~\ref{fig:l_mod} we show the fundamental mode for dipolar scalar perturbations ($\ell=1$). Just as in the monopolar case, the dipolar also has a critical mass (represented by the vertical dotted line) below which unstable modes appear. The critical mass for the $\ell=1$ is well below the one for $\ell=0$, so that the BH is already quite unstable in that region. Note that, differently from the monopolar case, the dipolar fundamental mode does not become unstable in the transition.

We note that both the methods used here to compute the modes agree in the qualitative behavior. The CF method is more accurate to describe modes with higher-imaginary part when compared to the DI one. However, it gets increasingly difficult to compute the modes near the critical mass, as the real part of the modes is very small. This means that the numerical infinity for the DI should be relatively large to capture the wave behavior of the solutions. For the CF method, we have to consider terms up to $N\sim2000$ to obtain stable numerical results. Nonetheless, both methods agree with the critical point in a very precise way and they also have excellent agreement to compute the unstable purely imaginary modes.

In the next section we further investigate implications of the instability in dynamical processes, by looking into the scattering of wave packets by the Schwarzschild spacetime in sGB gravity. 

\subsection{Dynamical response to scalar pulses}

To illustrate the instability of the Schwarzschild spacetime, here we investigate the outcome of the interaction between the BH and a Gaussian packet. We use the procedure presented by Ref.~[\refcite{Gundlach:1993tp}].

In order to perform the integrations, we transform Eq.~\eqref{eq:eom_perturbation} by considering the null coordinates $(u,v)=(t-r_{\ast},t+r_{\ast})$. The equation takes the simple form
\begin{equation}
\left(\frac{\partial^2}{\partial u \partial v}+\frac{1}{4}V_{\rm eff}\right)\varphi_1(u,v)=0.
\end{equation}
We solve the above equation subjected to the following initial condition
\begin{equation}
\varphi_1(0,v)=A \exp \left(-\frac{(v-v_c)^2}{2\sigma}\right),~~~~
\varphi_1(u,0)=0.
\label{eq:init_cond}
\end{equation}
For this work, we consider $A=1$, $v_c=20$, and $\sigma=1$. It has been noted by many works that the generic response to an initial perturbation can be divided in three different stages~\cite{Berti:2009kk}. First, there is a prompt initial signal, which is generically data-dependent. Second, a ringdown stage at intermediate times, which depends on the interaction between the BH and the wave, oscillating and decaying according to the characteristic QNMs of the BH. Finally, the late-time behavior of the perturbations. If the field is massless, the late-time behavior depends essentially on the multipolar perturbation, being a power-law decay~\cite{Gundlach:1993tp}. Besides changing the power-law behavior of late-time signals, massive fields induce some oscillations on this stages, related to the quasi-bound modes of the field~\cite{Burko:2004jn,Koyama:2001ee,Koyama:2001qw}. The late-time behavior of massive fields is also present when there are couplings between massive and massless fields~\cite{Macedo:2018txb}. 

\begin{figure}
	\includegraphics[width=\linewidth]{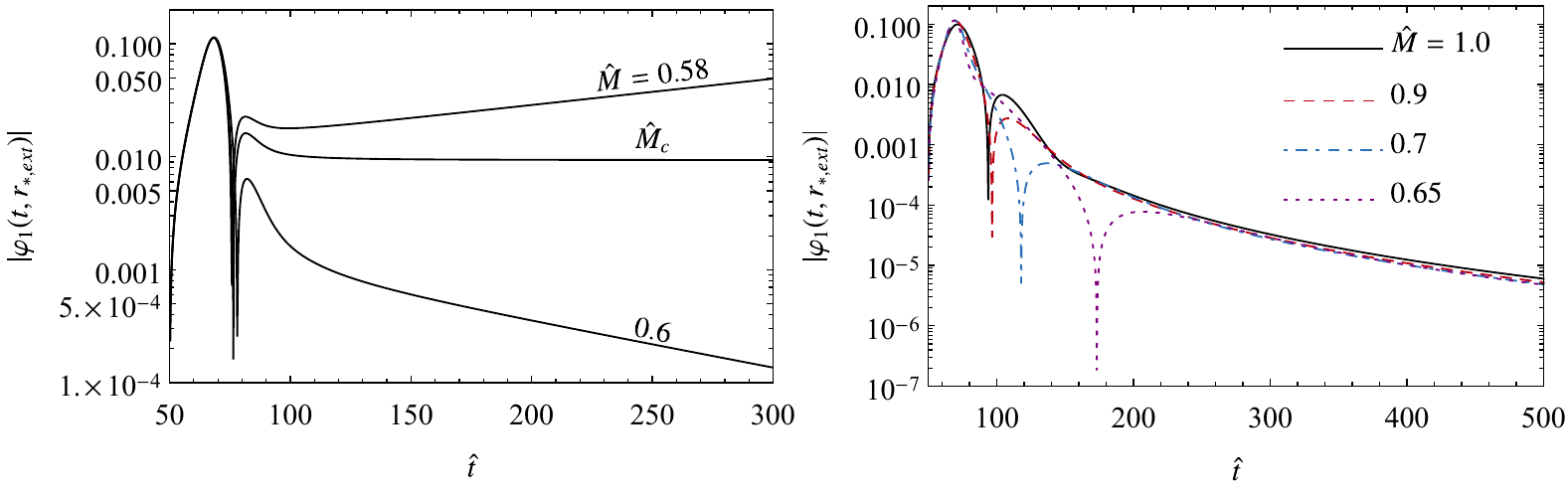}
	\caption{Time-evolution for scalar waves, $\ell=0$, with initial condition given by Eqs.~\eqref{eq:init_cond}. \textit{Left panel.} We see that for $\hat M <\hat M_c\approx 0.587$ the perturbation grows exponentially at late-times, representing the instability of the Schwarzschild BH. \textit{Right panel.} The signal initially oscillates, with frequency that depends on the value of $\hat M$, decaying and at late-times the behavior is universal (power-law). Note that the case $\hat M=1$ is already quite similar to the one of scalar fields in GR.}
	\label{fig:l0_time}
\end{figure}

\begin{figure}
	\includegraphics[width=\linewidth]{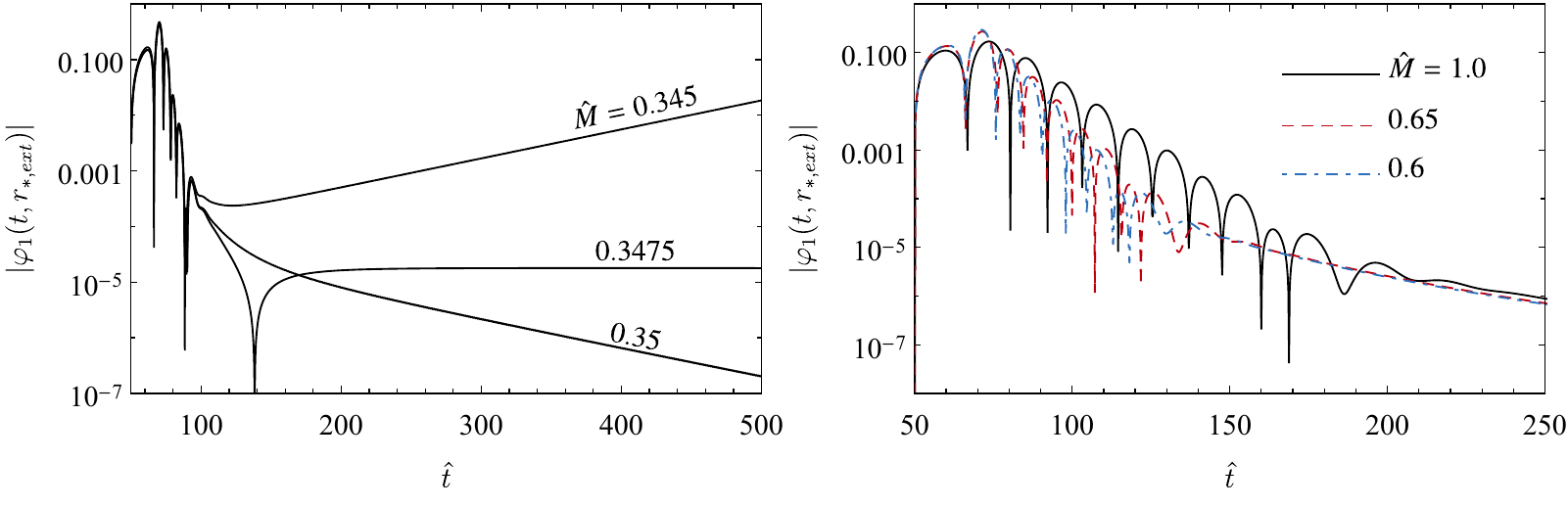}
	\caption{Same as Fig.~\ref{fig:l0_time}, but considering $\ell=1$ and different intervals for $\hat M$. Note that the onset of the instability for the $\ell=1$ is different from the $\ell=0$, being smaller.}
	\label{fig:l1_time}
\end{figure}

As mentioned above, in the ringdown stage the signal oscillates and decays according to the QNMs of the BH. By analyzing this stage we can, therefore, obtain these frequencies. This can be done by extracting the wave-function at a fixed point far from the BH---here we consider $r_{\ast , ext} =50$---fitting the signal to a superposition of waves in the form
\be
\varphi_1(t,r_{\ast , ext})\sim \sum_i A_{i} e^{-i \omega_i t},
\ee
where $A_i$ and  $\omega_i$ are the obtained by fitting the above expression with the numerical evolution using, for instance, the Prony method (see, e.g., Ref.~[\refcite{Berti:2007dg}]). We present our numerical results below.

As mentioned before, the Schwarzschild BH is unstable in the regime $M<M_c$. In the time-evolution of the signal, this translates into a exponential growth of the perturbations at late-times. In Fig.~\ref{fig:l0_time} we show the time-evolution of an initial Gaussian [cf.~\eqref{eq:init_cond}], for different values of the mass $\hat M$. In the left panel of Fig.~\ref{fig:l0_time} we focus on the regime in which the instability occurs. At the onset, the scalar field goes to a constant value at late-times, meaning that the BH is developing a hairy configuration~\cite{Benkel:2016kcq}. Note that to fully explore the hairy solution one needs to go beyond the linearized equations.

In the right-panel of Fig.~\ref{fig:l0_time} we focus on a different regime, in which the Schwarzschild BH is always stable. We can see that the frequency (as measured by $\hat t$) seems to decrease as we decrease the mass of the BH. This is consistent with the mode analysis presented previously. In the plot we can also see that the universal power-law late-time behavior is still the same, as it should.

Going further into higher-multipoles, we can see that the onset of the instability changes to smaller values of $\hat M$, as mentioned previously. In Fig.~\ref{fig:l1_time} we plot the result of the time-evolution for the dipolar ($\ell=1$) mode. In the left-panel, once again, we focus on the regime in which the field starts to grow exponentially. Since the threshold in smaller than the one for $\ell=0$, the black hole is already (very) unstable in that regime, so we have an additional unstable mode. Note, however, that differently from the $\ell=0$ case, the frequency of the oscillation in the stable regime increases as $\hat M$ decreases (as measured by $\hat t$). Note that, differently from the $\ell=0$ case, the fundamental mode is always stable in the $\ell=1$ in the regime studied here, so that the instability does not comes from the fundamental mode.

\subsection{Critical BH mass and self-interactions}

The instability suffered by the scalar field has a tachyonic nature: The beyond-GR terms behave as an effective negative mass-squared~\cite{Silva:2017uqg}. Up to now, we considered that the scalar mass is zero, for simplicity. Our model, however, do consider a scalar field mass $\mu$, which we wish to understand the influence for the BH\footnote{We recall that we consider linear perturbations with no background field. In this way, the quartic term does not enter in the equations;.}. Since $\mu^2>0$ in general, the scalar field mass combined with the effective one changes the value threshold mass $\hat M_c$, increasing the mass-range for stable Schwarzschild BHs.

\begin{figure}
	\centering
	\includegraphics[width=0.49\textwidth]{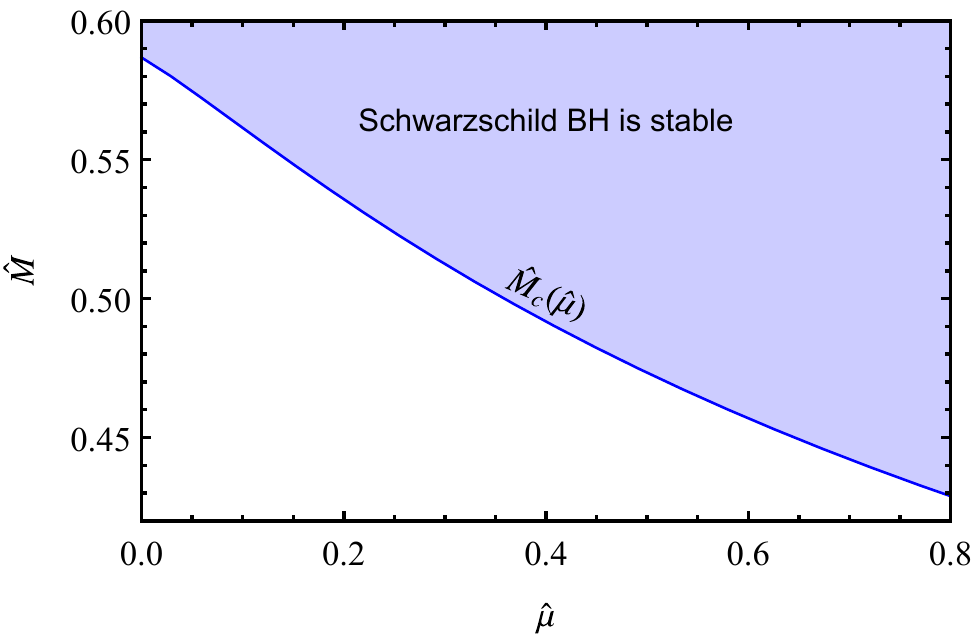}
	\caption{Critical mass $\hat M_c$ as function of the scalar field mass $\mu$. We see that $\hat M_c$ monotonically decreases with $\mu$, decreasing the range of unstable Schwarzschild BHs.}
	\label{fig:mass_m}
\end{figure}

In Fig.~\ref{fig:mass_m} we plot the value of the critical mass $\hat M_c$ as function of the scalar field mass. As predicted, the scalar field mass increase the range of stability for Schwarzschild BH, competing with the $\eta$ term in the effective potential~\eqref{eq:veff}. A similar result holds for higher overtones~\cite{Macedo:2019sem}, with the critical mass (line) being smaller.

\section{Nonlinear solutions and black hole scalarization}\label{sec:scal}

As mentioned before, the linear analysis tell us about the stability of the Schwarzschild BH, but does not show the outcome of the instability. As the scalar field grows, it is natural to ask ourselves whether it can balance itself against gravity to provide hair to the BH. In the next sections, we explore numerical solutions of the sGB gravity with non-trivial scalar fields and investigate their stability properties.

\subsection{Scalarized solutions and their stability}

\begin{figure}
	\centering
	\includegraphics[width=0.49\linewidth]{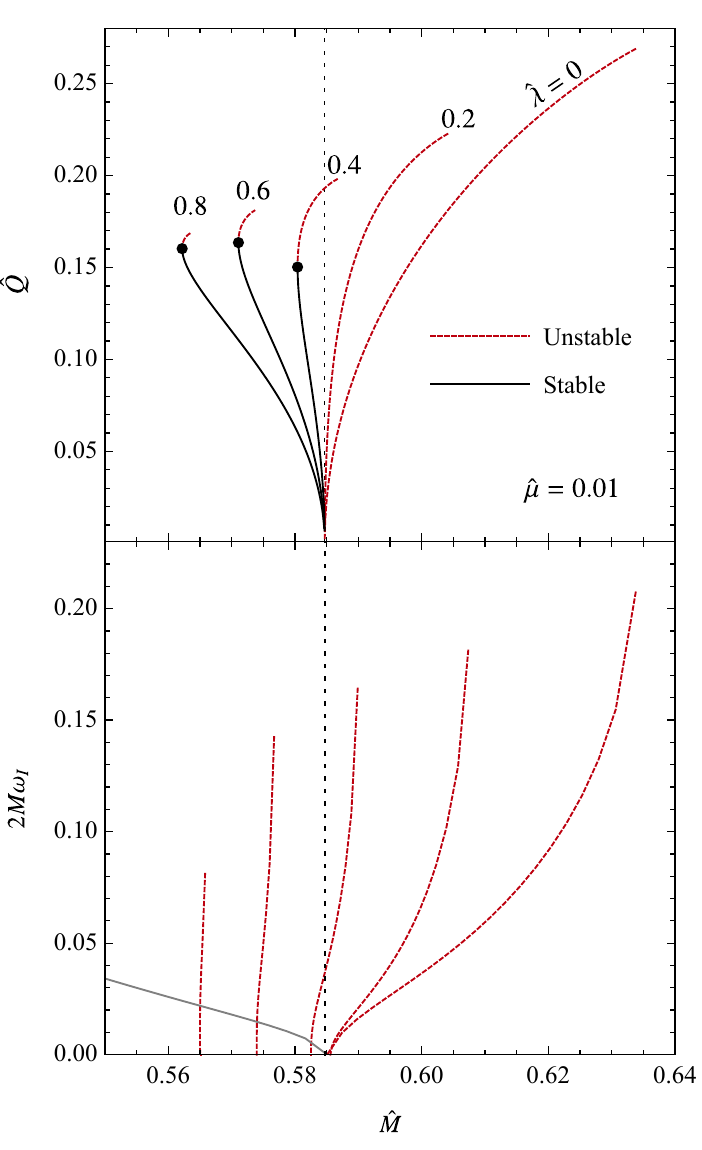}
	\includegraphics[width=0.49\linewidth]{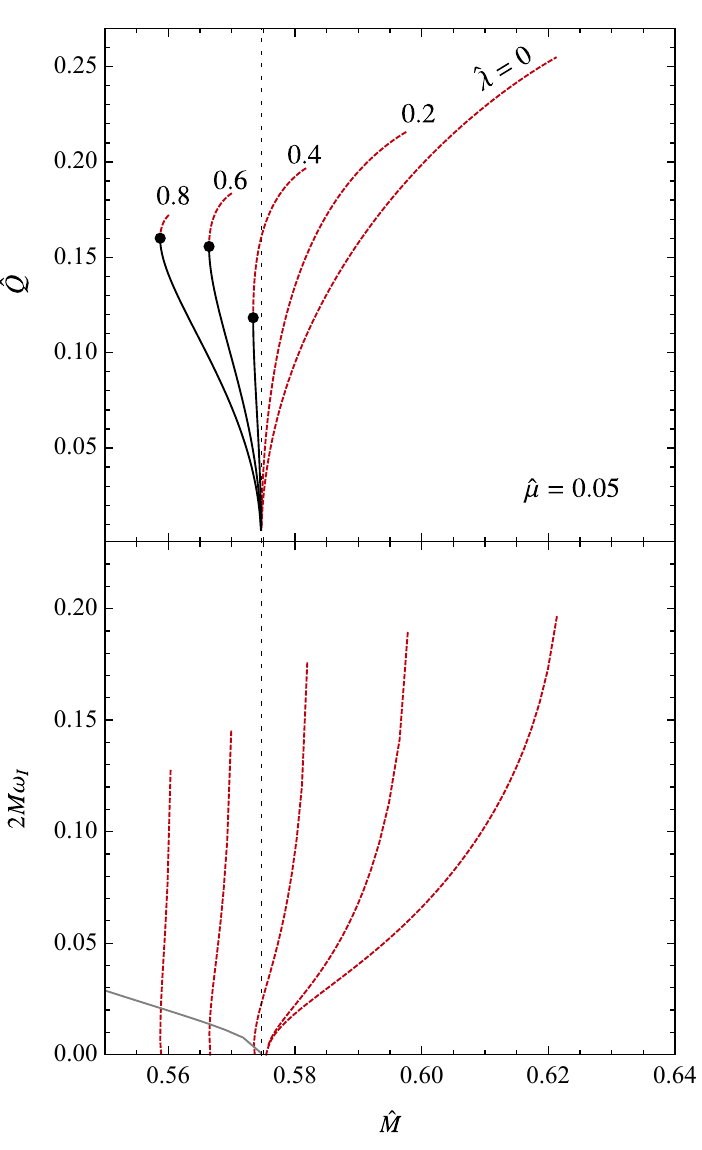}
	\caption{\textit{Upper panels.} Charge-mass diagram for the background scalarized solutions. The dashed red lines correspond to unstable configurations. The vertical dotted line marks the critical value of the mass $\hat M_c$. Note that the critical mass is different for each mass (compare left and right panels). As the quartic term in the self-interaction increases, the solution migrates to the stable side. \textit{Bottom panels.} The purely imaginary unstable modes for the scalarized solution (dashed curves) and the ones for Schwarzschild (gray solid line). Note that depending on $(\mu,\lambda)$, we can have \textit{two} scalarized solution for the same mass $\hat M$, with one being unstable.}
	\label{fig:scalarized}
\end{figure}

We can integrate the equations of motion \eqref{eq:eintt}--\eqref{eq:scalar_back}, subject to the boundary conditions~\eqref{eq:metric_cond}--\eqref{eq:phi_der} and~\eqref{eq:metric_inf}--\eqref{eq:scalar_inf}, searching for solutions with nontrivial scalar fields. Here we show a selection of the results, illustrating how the self-interactions and the coupling function changes stability properties of scalarized BHs. More details can be seem in Refs.~[\refcite{Silva:2018qhn,Macedo:2019sem}].

Let us first focus on the effects on self-interactions, setting $\zeta=0$ in the coupling function. In the effective-field theory picture, self-interactions are more important than higher-order couplings. In top panels of Fig.~\ref{fig:scalarized} we show charge-mass diagrams of the background solutions, considering different values for the self-interacting terms. Note the vertical dotted line---denoting the threshold for the instability---moves to the left when the mass $\hat \mu$ increases, in accordance with Fig.~\ref{fig:mass_m} (compare the left and right panels). Moving to the quartic term in the self-interaction, regulated by $\hat\lambda$, we see that increasing it makes the curves bend more to the left part of the diagrams, where the Schwarzschild solution is unstable.

Since the Schwarzschild spacetime is unstable in that region, it is natural to ask ourselves whether the scalarized solution can be stable. In this sense, we have that the Schwarzchild BH would spontaneouly grow hair, moving to a solution with non-trivial scalar fields. We perform the radial stability analysis by integrating the differential equation~\eqref{eq:per_scalar} for scalarized solutions, requiring the field to vanish at the boundaries (condition for instability\cite{Torii:1998gm,Blazquez-Salcedo:2016enn}). The result can be seen in the bottom panels of Fig.~\ref{fig:scalarized}, where we see indeed that there are stable scalarized solution in the region where the Schwarzschild one is unstable. The solutions represented dashed lines are radially unstable, while the ones represented by the solid line are stable. The instability also depends on the value of $\lambda$: Bigger the fourth-order potential, wider the range of masses $\hat M$ in which the scalarized solutions are stable. 

Note also from Fig.~\ref{fig:scalarized} that not all solutions in the left part of the diagram are stable. There exist values of masses $\hat M$ in which two scalarized solution exist, in addition to the Schwarzschild spacetime\footnote{Note that there are actually five solutions, considering  the shift symmetry in the scalar field.}. The radial stability analysis reveals that, in these cases, solutions with higher charges are actually unstable. It would be interesting to explore such scenarios---with three solutions with the same mass---in dynamical processes. We that all solutions are nodeless, so the higher $Q$ unstable solutions are still in the bound state configuration for the scalar field.

\begin{figure}
	\centering
	\includegraphics[width=0.5\linewidth]{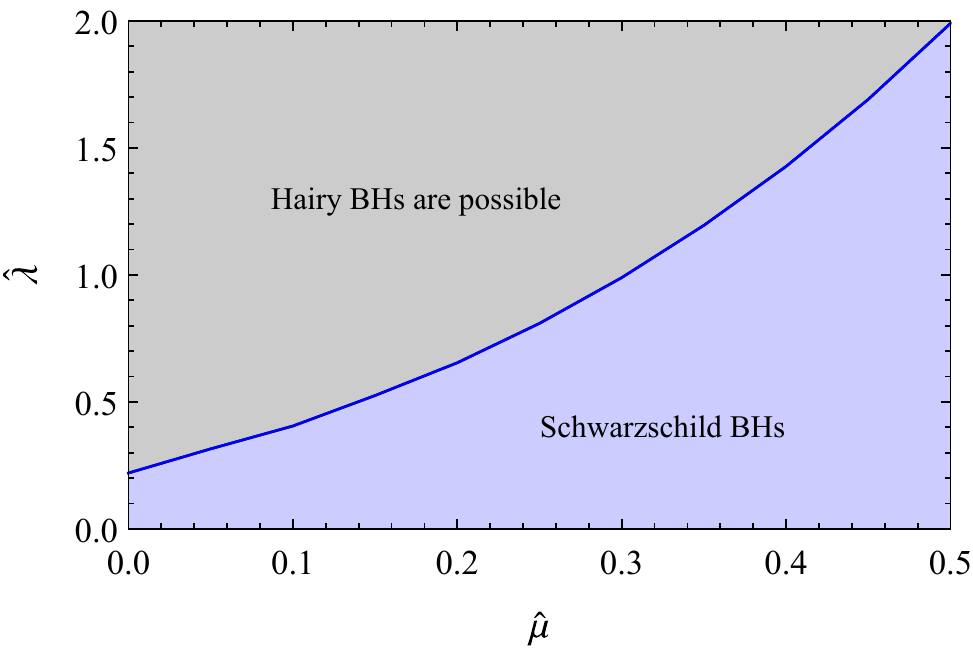}
	\caption{Diagram depicting possible outcomes in the gravitational collapse, depending on the self-interacting terms, considering $\zeta=0$. Above the line, depending on the mass of the progenitor, we conjecture that it is possible to form scalarized hairy BHs. Bellow the line, only the Schwarzschild spacetime is possible as a BH solution.}
	\label{fig:diag}
\end{figure}

We conclude that self-interactions alone are enough to stabilize scalarized BHs. There is a critical value of the pair $(\hat\mu,\hat\lambda)$ that this occurs. For instance, fixing $\hat\mu=0.01$, Fig.~\ref{fig:scalarized} shows that the critical $\hat\lambda$ to stabilize solutions is $\hat\lambda\gtrsim 0.2$. The analysis of the critical parameters allows us to draw a stability diagram considering self-interactions. In Fig.~\ref{fig:diag} we show the stability diagram. For values of $(\hat\mu,\hat\lambda)$ above the critical line, stable scalarized solution exist and can be, depending on the progenitor mass, possible endstate for the gravitational collapse. Bellow the diagram, all scalarized solutions are unstable and, therefore, would never form dynamically. Schwarzschild spacetime is the only possible endstate if a BH is to be form.

\begin{figure}
	\includegraphics[width=\linewidth]{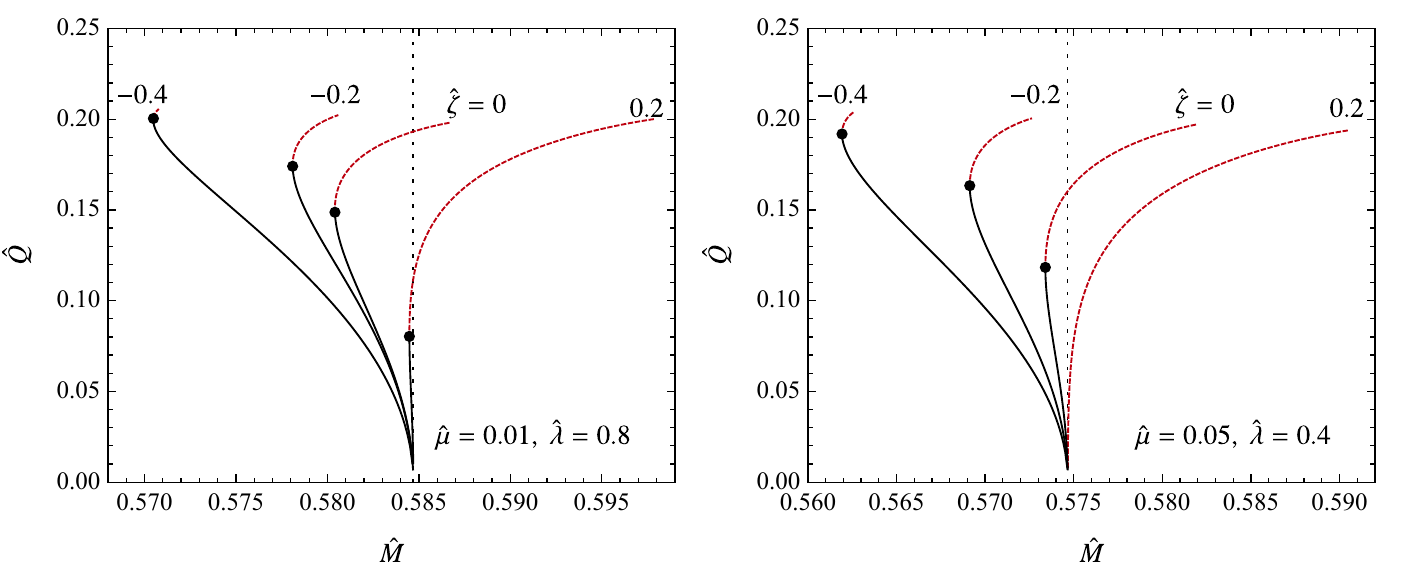}
	\caption{Charge-mass diagrams for scalarized solutions, considering the combined effects of the quartic coupling and the self-interacting terms.}
	\label{fig:coupling}
\end{figure}

To complete the picture, it is also instructive to analyze the combined effects of self-interactions and nonlinear coupling functions. It was recently pointed out in a couple of works that nonlinearities can quench the instability of scalarized BHs~\cite{Minamitsuji:2018xde,Silva:2018qhn}. The quenching of the instability also explains why the exponential model adopted by Ref.~[\refcite{Doneva:2017bvd}]  is stable, while the quadratic adopted in Ref.~[\refcite{Silva:2017uqg}] is not\cite{Blazquez-Salcedo:2018jnn}. In Fig.~\ref{fig:coupling} we show the influence of $\hat\zeta$ in the charge-mass diagram of scalarized solutions, for fixed values of $(\hat\mu,\hat\lambda)$. Decreasing $\hat\zeta$ stabilizes the scalarized solutions. We note that, by comparing with the results of Ref.~[\refcite{Silva:2018qhn}], the existence of two scalarized solutions with the same mass (one stable and the other unstable) is a characteristic of the models with self-interactions.

\subsection{Null geodesic frequency and Lyapunov exponents}\label{sec:geodesic}

It is instructive to consider the null-geodesic quantities in the scalarized BHs and quantify how different they are from the GR case. The null-geodesics have an impact in the shadow of the BH, a quantity that can be measured by the Event Horizon Telescope~\cite{Akiyama:2019cqa}. Moreover, there is a close relation of null-geodesic quantities and the higher $\ell$ QNMs of BHs~\cite{Cardoso:2008bp}
\footnote{Note that, in general, the null-geodesic quantities of the metric are not always appropriate to describe the modes. See Refs.~[\refcite{Glampedakis:2019dqh,Silva:2019scu}] for details.}.

By computing null-geodesics quantities, one can show that the equations can be reduced to an effective balance energy equation. The procedure can be seen in Ref.~[\refcite{Cardoso:2008bp}]. The frequency of the circular null unstable geodesic (light-ring) is given by
\begin{equation}
	\Omega_l=\frac{A(r_l)^{1/2}}{r_l},
\end{equation}
where $r_l$ is the radius of the orbit, which satisfies
\be
2 A(r_l)-r_l A'(r_l)=0.
\ee
When the QNM geodesic correspondence is valid, the light-ring frequency is closely related to the real part of the modes. We also note that the light-ring frequency is related to the critical impact parameter through $b_c=\Omega_l^{-1}$, a quantity that is relevant in computing the shadow of the BH. 

The instability timescale of the circular null-geodesic is related to the Lyapunov exponent, given by
\begin{equation}
	\Lambda =\left.\sqrt{ \frac{-r^2}{2A(r)}\left(\frac{d^2}{dr_\ast^2}\frac{A(r)}{r^2}\right)}\right|_{r=r_l}.
\end{equation}
In the QNM geodesic correspondence, the Lyapunov exponent is related to the imaginary part of the modes.

\begin{figure}
	\centering
	\includegraphics[width=0.49\textwidth]{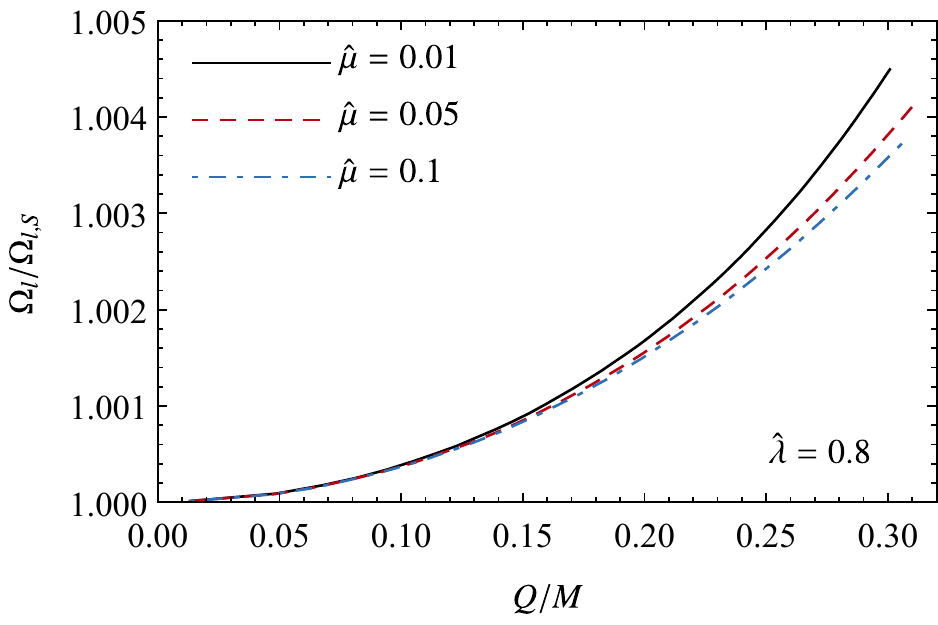}
	\includegraphics[width=0.49\textwidth]{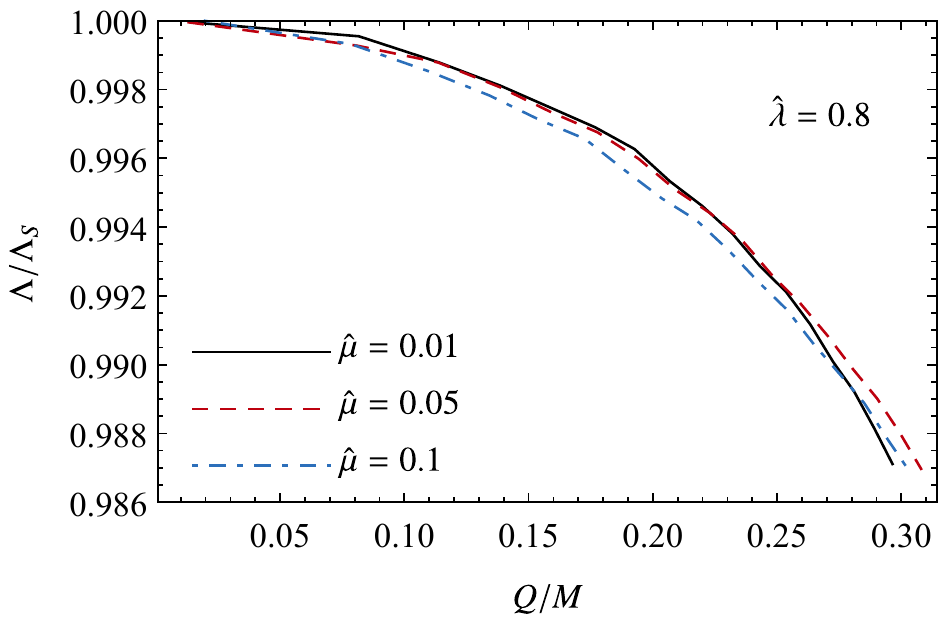}
	\caption{Light-ring frequency and Lyapunov exponent, normalized by the GR quantities.}
	\label{fig:geo}
\end{figure}

In Fig.~\ref{fig:geo} we show the light-ring frequency and the Lyapunov exponent of scalarized BHs for some values of $(\hat\mu,\hat\lambda)$, normalized by the GR value. For the plots we consider only the quadratic coupling function. As we move to higher-charge configurations, we can see that the light-ring frequency increases, while the Lyapunov exponent decreases. Note, however, that the changes are very small, being smaller than $1\%$ for the frequency and smaller than $2\%$ for the Lyapunov exponent. This is natural to expect, as changes in the geometry of the BH at the light-ring, roughly $3M$, are already quite small: The GB term for Schwarzschild at this location is ${\cal G}M^4\approx 0.066$. The deviations indicate that the QNM of scalarized BHs can be very similar to the GR ones.\footnote{Note, however, that when one considers nonlinear couplings together with rotation the result can change considerably. See Ref.~[\refcite{Cunha:2019dwb}].}

\section{Final remarks}\label{sec:final}

Theories that present spontaneous scalarized BHs are interesting strawman to consider deviations of GR. Since they possess the same GR solutions, they are basically identical to GR in the kinematic scenario, such as the ones considered by electromagnetic computations. In this work, we shown some features of Schwarzschild BHs in sGB, highlighting the differences from the GR case. We also overview scalarized solutions and compute the null-geodesic quantities for hairy BHs.

There are many possible venues to extend the work presented here. The mass-range stability should be further scrutinized, to understand how formation of BHs evolve in those regimes. For instance, the analysis presented here predicts that no bald BH should form in the regime $\hat M <\hat M_c$. To access that, collapse models in extension of GR are necessary, which is an area still in development. Another point is that our stability analysis are mostly valid for radially symmetric perturbations. Nonradial perturbations is an important extension that will be tackle by future works. Finally, it would be interesting to consider radiation emission processes in these theories, to see whether the additional polarization modes appear in the GW signals~\cite{Blazquez-Salcedo:2016enn}.

\section*{Acknowledgments}
The author would like to thank Emanuele Berti, Leonardo Gualtieri, Jeremy Sakstein, Hector O. Silva, and Thomas Sotiriou for fruitful discussions. The author also thanks Conselho Nacional de Desenvolvimento Cient\'ifico e Tecnol\'ogico (CNPq) and Coordena\c{c}\~ao de Aperfei\c{c}oamento de Pessoal de N\'ivel Superior (CAPES), from Brazil, for partial financial support.

\bibliography{biblio}

\begin{thebibliography}{10}

\bibitem{Abbott:2016blz}
 LIGO Scientific, Virgo Collaboration (B.~P. Abbott {\em et~al.}), {\em Phys.
  Rev. Lett.} {\bf 116}  (2016)   061102,
  \href{http://arxiv.org/abs/1602.03837}{{\ttfamily arXiv:1602.03837 [gr-qc]}}.

\bibitem{Abbott:2018lct}
 LIGO Scientific, Virgo Collaboration (B.~P. Abbott {\em et~al.})  (2018)
  \href{http://arxiv.org/abs/1811.00364}{{\ttfamily arXiv:1811.00364 [gr-qc]}}.

\bibitem{Sakstein:2017xjx}
J.~Sakstein and B.~Jain, {\em Phys. Rev. Lett.} {\bf 119}  (2017)   251303,
  \href{http://arxiv.org/abs/1710.05893}{{\ttfamily arXiv:1710.05893
  [astro-ph.CO]}}.

\bibitem{Ezquiaga:2017ekz}
J.~M. Ezquiaga and M.~Zumalac\'arregui, {\em Phys. Rev. Lett.} {\bf 119}
  (2017)   251304, \href{http://arxiv.org/abs/1710.05901}{{\ttfamily
  arXiv:1710.05901 [astro-ph.CO]}}.

\bibitem{Creminelli:2017sry}
P.~Creminelli and F.~Vernizzi, {\em Phys. Rev. Lett.} {\bf 119}  (2017)
  251302, \href{http://arxiv.org/abs/1710.05877}{{\ttfamily arXiv:1710.05877
  [astro-ph.CO]}}.

\bibitem{Baker:2017hug}
T.~Baker, E.~Bellini, P.~G. Ferreira, M.~Lagos, J.~Noller and I.~Sawicki, {\em
  Phys. Rev. Lett.} {\bf 119}  (2017)   251301,
  \href{http://arxiv.org/abs/1710.06394}{{\ttfamily arXiv:1710.06394
  [astro-ph.CO]}}.

\bibitem{Dima:2017pwp}
A.~Dima and F.~Vernizzi, {\em Phys. Rev.} {\bf D97}  (2018)   101302,
  \href{http://arxiv.org/abs/1712.04731}{{\ttfamily arXiv:1712.04731 [gr-qc]}}.

\bibitem{Crisostomi:2017lbg}
M.~Crisostomi and K.~Koyama, {\em Phys. Rev.} {\bf D97}  (2018)   021301,
  \href{http://arxiv.org/abs/1711.06661}{{\ttfamily arXiv:1711.06661
  [astro-ph.CO]}}.

\bibitem{Langlois:2017dyl}
D.~Langlois, R.~Saito, D.~Yamauchi and K.~Noui, {\em Phys. Rev.} {\bf D97}
  (2018)   061501, \href{http://arxiv.org/abs/1711.07403}{{\ttfamily
  arXiv:1711.07403 [gr-qc]}}.

\bibitem{Berti:2015itd}
E.~Berti {\em et~al.}, {\em Class. Quant. Grav.} {\bf 32}  (2015)   243001,
  \href{http://arxiv.org/abs/1501.07274}{{\ttfamily arXiv:1501.07274 [gr-qc]}}.

\bibitem{Sotiriou:2015lxa}
T.~P. Sotiriou, {\em Lect. Notes Phys.} {\bf 892}  (2015) 3,
  \href{http://arxiv.org/abs/1404.2955}{{\ttfamily arXiv:1404.2955 [gr-qc]}}.

\bibitem{Herdeiro:2015waa}
C.~A.~R. Herdeiro and E.~Radu, {\em Int. J. Mod. Phys.} {\bf D24}  (2015)
  1542014, \href{http://arxiv.org/abs/1504.08209}{{\ttfamily arXiv:1504.08209
  [gr-qc]}}.

\bibitem{Silva:2017uqg}
H.~O. Silva, J.~Sakstein, L.~Gualtieri, T.~P. Sotiriou and E.~Berti, {\em Phys.
  Rev. Lett.} {\bf 120}  (2018)   131104,
  \href{http://arxiv.org/abs/1711.02080}{{\ttfamily arXiv:1711.02080 [gr-qc]}}.

\bibitem{Doneva:2017bvd}
D.~D. Doneva and S.~S. Yazadjiev, {\em Phys. Rev. Lett.} {\bf 120}  (2018)
  131103, \href{http://arxiv.org/abs/1711.01187}{{\ttfamily arXiv:1711.01187
  [gr-qc]}}.

\bibitem{Antoniou:2017acq}
G.~Antoniou, A.~Bakopoulos and P.~Kanti, {\em Phys. Rev. Lett.} {\bf 120}
  (2018)   131102, \href{http://arxiv.org/abs/1711.03390}{{\ttfamily
  arXiv:1711.03390 [hep-th]}}.

\bibitem{Damour:1993hw}
T.~Damour and G.~Esposito-Far\`ese, {\em Phys. Rev. Lett.} {\bf 70}  (1993)
  2220.

\bibitem{EspositoFarese:2004cc}
G.~Esposito-Farese, {\em AIP Conf. Proc.} {\bf 736}  (2004) 35,
  \href{http://arxiv.org/abs/gr-qc/0409081}{{\ttfamily arXiv:gr-qc/0409081
  [gr-qc]}}.

\bibitem{Macedo:2019sem}
C.~F.~B. Macedo, J.~Sakstein, E.~Berti, L.~Gualtieri, H.~O. Silva and T.~P.
  Sotiriou, {\em Phys. Rev.} {\bf D99}  (2019)   104041,
  \href{http://arxiv.org/abs/1903.06784}{{\ttfamily arXiv:1903.06784 [gr-qc]}}.

\bibitem{Silva:2018qhn}
H.~O. Silva, C.~F.~B. Macedo, T.~P. Sotiriou, L.~Gualtieri, J.~Sakstein and
  E.~Berti, {\em Phys. Rev.} {\bf D99}  (2019)   064011,
  \href{http://arxiv.org/abs/1812.05590}{{\ttfamily arXiv:1812.05590 [gr-qc]}}.

\bibitem{Cardoso:2008bp}
V.~Cardoso, A.~S. Miranda, E.~Berti, H.~Witek and V.~T. Zanchin, {\em Phys.
  Rev.} {\bf D79}  (2009)   064016,
  \href{http://arxiv.org/abs/0812.1806}{{\ttfamily arXiv:0812.1806 [hep-th]}}.

\bibitem{Will:2014kxa}
C.~M. Will, {\em Living Rev. Rel.} {\bf 17}  (2014)  ~4,
  \href{http://arxiv.org/abs/1403.7377}{{\ttfamily arXiv:1403.7377 [gr-qc]}}.

\bibitem{Torii:1998gm}
T.~Torii and K.-i. Maeda, {\em Phys. Rev.} {\bf D58}  (1998)   084004.

\bibitem{Kanti:1995vq}
P.~Kanti, N.~E. Mavromatos, J.~Rizos, K.~Tamvakis and E.~Winstanley, {\em Phys.
  Rev.} {\bf D54}  (1996) 5049,
  \href{http://arxiv.org/abs/hep-th/9511071}{{\ttfamily arXiv:hep-th/9511071
  [hep-th]}}.

\bibitem{Blazquez-Salcedo:2018jnn}
J.~L. Bl\'azquez-Salcedo, D.~D. Doneva, J.~Kunz and S.~S. Yazadjiev, {\em Phys.
  Rev.} {\bf D98}  (2018)   084011,
  \href{http://arxiv.org/abs/1805.05755}{{\ttfamily arXiv:1805.05755 [gr-qc]}}.

\bibitem{Nollert:1999ji}
H.-P. Nollert, {\em Class. Quant. Grav.} {\bf 16}  (1999) R159.

\bibitem{Berti:2007dg}
E.~Berti, V.~Cardoso, J.~A. Gonzalez and U.~Sperhake, {\em Phys. Rev.} {\bf
  D75}  (2007)   124017, \href{http://arxiv.org/abs/gr-qc/0701086}{{\ttfamily
  arXiv:gr-qc/0701086 [gr-qc]}}.

\bibitem{Konoplya:2011qq}
R.~A. Konoplya and A.~Zhidenko, {\em Rev. Mod. Phys.} {\bf 83}  (2011) 793,
  \href{http://arxiv.org/abs/1102.4014}{{\ttfamily arXiv:1102.4014 [gr-qc]}}.

\bibitem{Pani:2013pma}
P.~Pani, {\em Int. J. Mod. Phys.} {\bf A28}  (2013)   1340018,
  \href{http://arxiv.org/abs/1305.6759}{{\ttfamily arXiv:1305.6759 [gr-qc]}}.

\bibitem{Macedo:2016wgh}
C.~F.~B. Macedo, V.~Cardoso, L.~C.~B. Crispino and P.~Pani, {\em Phys. Rev.}
  {\bf D93}  (2016)   064053, \href{http://arxiv.org/abs/1603.02095}{{\ttfamily
  arXiv:1603.02095 [gr-qc]}}.

\bibitem{Chandrasekhar:1975zza}
S.~Chandrasekhar and S.~L. Detweiler, {\em Proc. Roy. Soc. Lond.} {\bf A344}
  (1975) 441.

\bibitem{Leaver:1985ax}
E.~W. Leaver, {\em Proc. Roy. Soc. Lond.} {\bf A402}  (1985) 285.

\bibitem{Nollert:1993zz}
H.-P. Nollert, {\em Phys. Rev.} {\bf D47}  (1993) 5253.

\bibitem{Silva:2014fca}
H.~O. Silva, C.~F.~B. Macedo, E.~Berti and L.~C.~B. Crispino, {\em Class.
  Quant. Grav.} {\bf 32}  (2015)   145008,
  \href{http://arxiv.org/abs/1411.6286}{{\ttfamily arXiv:1411.6286 [gr-qc]}}.

\bibitem{Benkel:2016kcq}
R.~Benkel, T.~P. Sotiriou and H.~Witek, {\em Phys. Rev.} {\bf D94}  (2016)
  121503, \href{http://arxiv.org/abs/1612.08184}{{\ttfamily arXiv:1612.08184
  [gr-qc]}}.

\bibitem{Gundlach:1993tp}
C.~Gundlach, R.~H. Price and J.~Pullin, {\em Phys. Rev.} {\bf D49}  (1994) 883,
  \href{http://arxiv.org/abs/gr-qc/9307009}{{\ttfamily arXiv:gr-qc/9307009
  [gr-qc]}}.

\bibitem{Berti:2009kk}
E.~Berti, V.~Cardoso and A.~O. Starinets, {\em Class. Quant. Grav.} {\bf 26}
  (2009)   163001, \href{http://arxiv.org/abs/0905.2975}{{\ttfamily
  arXiv:0905.2975 [gr-qc]}}.

\bibitem{Burko:2004jn}
L.~M. Burko and G.~Khanna, {\em Phys. Rev.} {\bf D70}  (2004)   044018,
  \href{http://arxiv.org/abs/gr-qc/0403018}{{\ttfamily arXiv:gr-qc/0403018
  [gr-qc]}}.

\bibitem{Koyama:2001ee}
H.~Koyama and A.~Tomimatsu, {\em Phys. Rev.} {\bf D64}  (2001)   044014,
  \href{http://arxiv.org/abs/gr-qc/0103086}{{\ttfamily arXiv:gr-qc/0103086
  [gr-qc]}}.

\bibitem{Koyama:2001qw}
H.~Koyama and A.~Tomimatsu, {\em Phys. Rev.} {\bf D65}  (2002)   084031,
  \href{http://arxiv.org/abs/gr-qc/0112075}{{\ttfamily arXiv:gr-qc/0112075
  [gr-qc]}}.

\bibitem{Macedo:2018txb}
C.~F.~B. Macedo, {\em Phys. Rev.} {\bf D98}  (2018)   084054,
  \href{http://arxiv.org/abs/1809.08691}{{\ttfamily arXiv:1809.08691 [gr-qc]}}.

\bibitem{Blazquez-Salcedo:2016enn}
J.~L. Bl\'azquez-Salcedo, C.~F.~B. Macedo, V.~Cardoso, V.~Ferrari,
  L.~Gualtieri, F.~S. Khoo, J.~Kunz and P.~Pani, {\em Phys. Rev.} {\bf D94}
  (2016)   104024, \href{http://arxiv.org/abs/1609.01286}{{\ttfamily
  arXiv:1609.01286 [gr-qc]}}.

\bibitem{Minamitsuji:2018xde}
M.~Minamitsuji and T.~Ikeda, {\em Phys. Rev.} {\bf D99}  (2019)   044017,
  \href{http://arxiv.org/abs/1812.03551}{{\ttfamily arXiv:1812.03551 [gr-qc]}}.

\bibitem{Akiyama:2019cqa}
 Event Horizon Telescope Collaboration (K.~Akiyama {\em et~al.}), {\em
  Astrophys. J.} {\bf 875}  (2019)  ~L1,
  \href{http://arxiv.org/abs/1906.11238}{{\ttfamily arXiv:1906.11238
  [astro-ph.GA]}}.

\bibitem{Glampedakis:2019dqh}
K.~Glampedakis and H.~O. Silva, {\em Phys. Rev.} {\bf D100}  (2019)   044040,
  \href{http://arxiv.org/abs/1906.05455}{{\ttfamily arXiv:1906.05455 [gr-qc]}}.

\bibitem{Silva:2019scu}
H.~O. Silva and K.~Glampedakis  (2019)
  \href{http://arxiv.org/abs/1912.09286}{{\ttfamily arXiv:1912.09286 [gr-qc]}}.

\bibitem{Cunha:2019dwb}
P.~V.~P. Cunha, C.~A.~R. Herdeiro and E.~Radu, {\em Phys. Rev. Lett.} {\bf 123}
   (2019)   011101, \href{http://arxiv.org/abs/1904.09997}{{\ttfamily
  arXiv:1904.09997 [gr-qc]}}.

\end{thebibliography}
\bibliographystyle{ws-ijmpd}

\end{document}